\documentclass[aps,pra,superscriptaddress,
 amsmath,amssymb,
 reprint,
]{revtex4-1}

\usepackage{dcolumn}
\usepackage{bm}
\usepackage{url}
\usepackage{pgfplots}
\usetikzlibrary{decorations.pathreplacing}
\pgfplotsset{compat=1.5}
\usepackage[export]{adjustbox}

\begin{document}
\title[]{ Spatial modulation of Joule losses to increase the normal zone propagation velocity in (RE)BaCuO tapes} 

\author{Jean-Hughes Fournier-Lupien}
\email{j-h.fournier@polymtl.ca}
\affiliation{Polytechnique Montr\'eal, Montr\'eal, QC, Canada, H3C 3A7}%
\affiliation{Karlsruhe Institute of Technology, Eggenstein-Leopoldshafen, D-76344 Germany}

\author{Christian Lacroix}%
\email{christian.lacroix@polymtl.ca}
\affiliation{Polytechnique Montr\'eal, Montr\'eal, QC, Canada, H3C 3A7}%

\author{Sebastian Hellmann}
\email{sebastian.hellmann@kit.edu}
\affiliation{Karlsruhe Institute of Technology, Eggenstein-Leopoldshafen, D-76344 Germany}

\author{Fr\'ed\'eric Sirois}
\email{f.sirois@polymtl.ca}
\affiliation{Polytechnique Montr\'eal, Montr\'eal, QC, Canada, H3C 3A7}%
\affiliation{Karlsruhe Institute of Technology, Eggenstein-Leopoldshafen, D-76344 Germany}
\date{\today}

\begin{abstract}
This paper presents a simple approach to increase the normal zone propagation velocity
in (RE)BaCuO thin films grown on a flexible metallic substrate, also called superconducting tapes.
The key idea behind this approach is to use a specific geometry of the silver thermal stabilizer that surrounds the superconducting tape. More specifically, a very thin layer of silver stabilizer is deposited on top of the superconductor layer, typically less than $100$~nm, while the remaining stabilizer (still silver) is deposited on the substrate side. Normal zone propagation velocities up to $170$~cm/s have been measured experimentally, corresponding to a stabilizer thickness of 20~nm on top of the superconductor layer. This is one order of magnitude faster than the speed measured on actual commercial tapes.
Our results clearly demonstrate that a very thin stabilizer on top of the superconductor layer leads to high normal zone propagation velocities. The experimental values are in good agreement with predictions realized by finite element simulations. Furthermore, the propagation of the normal zone during the quench was recorded in situ and in real time using a high-speed camera. Due to high Joule losses generated on both edges of the tape sample,
a ``U-shaped'' profile could be observed at the boundaries between the superconducting and the normal zones, which matches very closely the profile predicted by the simulations.
\end{abstract}

\maketitle
\section{Introduction}
Protection against destructive hot spots is one of the major unresolved issues for the implementation of electric power devices or electromagnets based on Second-Generation High-Temperature Superconductor Coated Conductors (2G HTS CC), also called HTS tapes. Hot spots arise when the current is close to the average critical current $I_c$ of the tape. Indeed, due to imperfections in the microstructure, $I_c$ varies unintentionally along the length of the tape \cite{Gurevich2001, Furtner2004, Kudymow2007, Colangelo2012a}, and when the transport current $I$ gets higher than the local $I_c$, this portion of the tape quenches (i.e. superconductivity is lost and a so-called \emph{normal zone} (NZ) appears). As a consequence, the current is rerouted into the stabilizer and the substrate, i.e. the metallic layers of the tape that are electrically conducting and which act as a thermal reservoir. This produces intense heating, resulting in a rapid local temperature rise (the hot spot) and the possible destruction of the tape if no appropriate action is taken.

One of the solutions that has been substantially investigated in the recent years \cite{Lacroix2013a, Lacroix2014, Lacroix2014a,Levin2010} consists in increasing the \emph{normal zone propagation velocity} (NZPV) of the tape, i.e.~the velocity at which the normal zone expands when a hot spot occurs. Low temperature superconductors (LTS) such as Nb$_3$Sn and NbTi have NZPVs in the range of 100-1000 cm/s \cite{Duckworth2001}, which
is fast enough to allow rapid quench detection and protective actions that avoid irreversible damages to the tape.
However, for standard coated conductor tapes made of HTS materials like (RE)BaCuO, the NZPV is much lower, i.e. $0.1 - 10$ cm/s \cite{Grabovickic2003, Wang2005a, Armenio2008, Wang2009, Park2010, Lu2013}, mainly because of their higher heat capacity, which makes quench detection much more difficult to realize.
If one succeeds to increase the NZPV of HTS tapes to values comparable to that observed in LTS wires, the protection of HTS-based devices would become much easier \cite{Maeda2014}.

\begin{figure}[h!]
	\includegraphics[width=8cm]{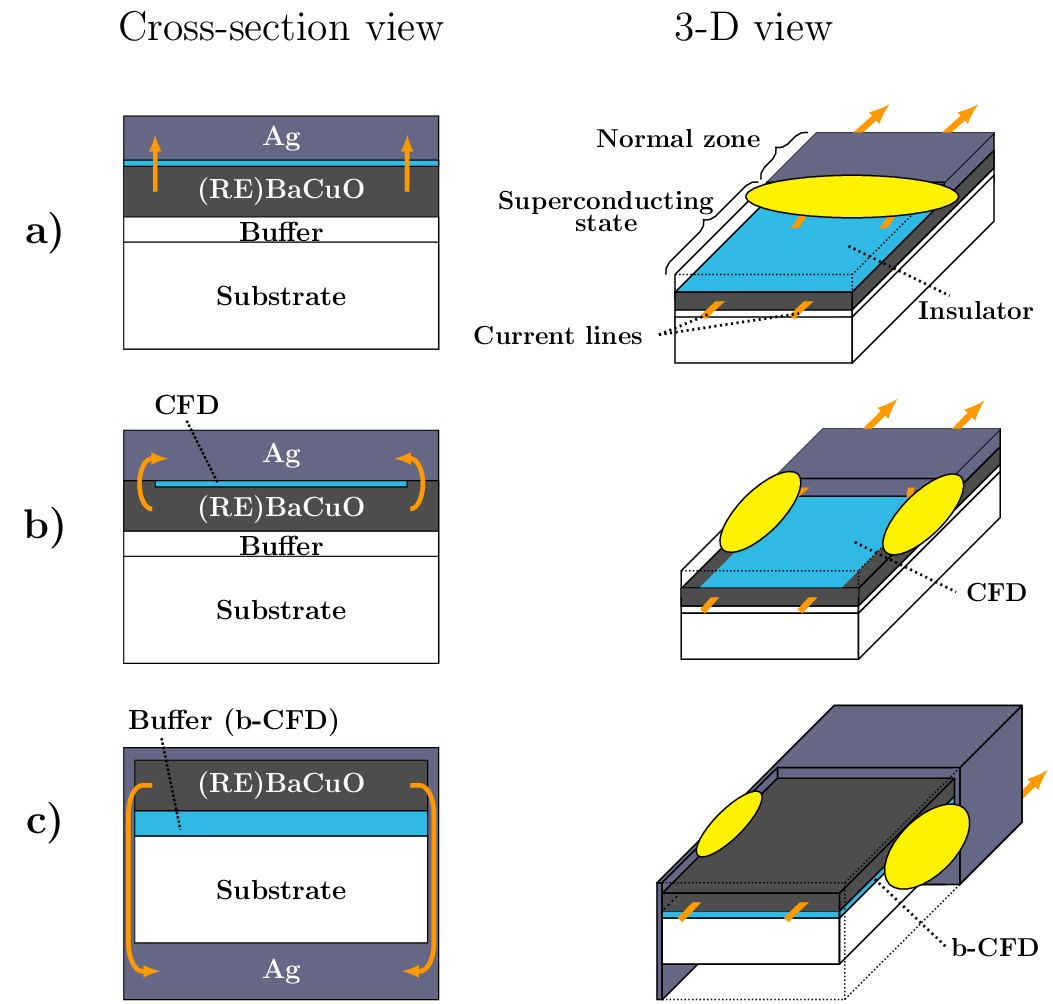}
     \caption{Cross-section views of different tape architectures proposed to increase the NZPV: a) the \emph{uniform architecture}, consisting of a uniform interfacial resistance at the Ag/(RE)BaCuO interface (cyan); b) the ``classical'' CFD architecture, consisting of a high contact resistance at the Ag/(RE)BaCuO interface (cyan), except at the edges of the tape, where the contact resistance is kept low; c) the ``b-CFD'' architecture, in which no high interfacial resistance is needed at the Ag/(RE)BaCuO interface, but which requires a thin top stabilizer and thick bottom stabilizer connected through a conducting edge metallization. The arrows show the current flow paths when the superconductor quenches.
             }
    \label{fig:1}
\end{figure}

It has been shown from numerical simulations
\cite{Levin2010, Chan2009} and experiments \cite{Lacroix2013a} that inserting a contact resistance (noted $R_i$, in $\mu\Omega.\textrm{cm}^2$) between the superconductor and the stabilizer of HTS tapes allows increasing their NZPV by more than one order of magnitude. In this case, the contact resistance is uniform across the tape width. We call this architecture, the ``uniform architecture'' (see Fig.~\ref{fig:1}a), which is very similar to that of commercially available tapes. It has been shown that the physical explanation of this phenomenon relies in the increase of the current transfer length (CTL)
associated with the increase of $R_i$ ~\cite{Levin2010,Lacroix2013a}. However, from a practical point of view, increasing $R_i$ has the drawback of significantly increasing the heat generation at the contacts for current injection, which is highly undesirable.

Recently, it has been proposed to use a patterned contact resistance at the Ag/(RE)BaCuO interface~\cite{Lacroix2014,Lacroix2014a}. If we pattern the contact resistance in order to have, for instance, a very high contact resistance in the middle of the tape and a low contact resistance on its edges, we obtain the so-called ``current flow diverter (CFD) architecture'' (see Fig.~\ref{fig:1}b)).
In this case, because of the presence of low resistance paths between the stabilizer and the superconductor, it is possible to inject current into the tape without burning the contacts.
As a result, a patterned contact resistance leads to faster expansion of the NZ for a given value of the global contact resistance. This can be explained partly by the current concentration occurring in the low contact resistance regions, which in turn increases the power density generated at the edges of the tape (yellow regions in Fig.~\ref{fig:1}b)). This partial quench across the width of the tape speeds up the normal zone propagation substantially.

Recent numerical simulations~\cite{Lacroix2017} showed that this solution is mostly effective for tapes with a thin stabilizer, a few micrometres at most. This means that for applications such as superconducting magnets, which require a thicker stabilizer (few tens of micrometres), a solution to increase the NZPV is still sought. Furthermore, from a practical point of view, patterning a contact resistance is a priori not straightforward to implement in an industrial fabrication process. Therefore, simpler means to implement the CFD concept are also sought.

In~\cite{Lacroix2017}, it has been proposed that the NZPV could be enhanced by simply modifying the stabilizer geometry without having to alter the Ag/(RE)BaCuO interface. This alternative approach has two major advantages. Firstly, the value of the contact resistance at the Ag/(RE)BaCuO interface can be kept as low as desired. Secondly, this approach remains effective even in the presence of a thick stabilizer.

In this paper, we demonstrate experimentally how to increase the NZPV by modifying the stabilizer geometry. We call this type of tape architecture ``b-CFD'' since it is now the buffer layers that divert the current flow when a normal zone appears, as illustrated in Fig.~\ref{fig:1}c). Extensive characterization of the NZPV was undertaken for different tape architecture, first by electrical measurements, and then by in situ video recording of a provoked quench in HTS tapes. Finally, we conducted numerical simulations that reproduced very well all experimental results, confirming our understanding of the underlying physics.

\section{Experimental Methodology}
\subsection{Samples preparation}
The HTS tape samples used in this work consist of 4 mm wide 2G HTS coated conductors, custom-made by Superconductor Technologies Inc (STI)\footnote{Superconductor Technologies Inc., (http://www.suptech.com/), Accessed November 11, 2016 }. The basic template for all samples, depicted in the Fig.~\ref{fig:1}a), consisted of a 0.8~$\mu$m thick Yttrium-Barium-Copper-Oxygen (YBa$_2$Cu$_3$O$_{7-\delta}$) layer grown on a 100~$\mu$m thick Hastelloy$^\text{\textregistered}$ substrate (C-276$^{\text{TM}}$) by reactive co-evaporation cyclic depositions and reactions (RCE-CDR) deposition method~\cite{Matias2009a}. Lattice matching between the (RE)BaCuO layer and the substrate was realized through a stack of buffer layers composed of CeO$_2$/MgO/[Y+Zr]$_2$O$_3$ (total thickness of $\approx$ 0.58 to 0.78~$\mu$m).

A 20 nm thick silver (Ag) capping layer has been deposited \textit{in-situ} on the (RE)BaCuO layer to ensure a low contact resistance at the Ag/(RE)BaCuO interface. Note that no oxygen annealing treatment has been done on the samples. Such a treatment, when applied on thin Ag layers (typically less than 500~nm) is known to compromise the integrity of the Ag layer by generating Ag aggregates (islands) when the temperature reaches a few hundred degrees (typically above $\approx400-500^{\circ}$C)~\cite{Ekin2006}.

A total of 4 samples (4 mm $\times$ 12 cm) with different metallic (Ag) stabilizer geometry were subsequently fabricated from the tape template described above. Then, a certain amount of stabilizer (still Ag) was added on the samples using RF sputtering deposition. Our setup allowed sputtering the desired thickness of Ag on the top (superconductor) side ($t_{\text{top}}$) and on the bottom (substrate) side ($t_{\text{bot}}$). We made sure that the total Ag thickness $t_\text{tot}=t_{\text{top}}+t_{\text{bot}}$ deposited on each sample was the same, i.e. $t_\text{tot}=~2~\mu$m. The exact Ag thicknesses on each side of the tape are given in the table~\ref{tab:bCFD_samples}. Special care has been taken to sputter Ag on both edges of the tape in order to electrically connect the top and bottom Ag layers. Furthermore, critical current measurements ($T=77$~K) carried out on each sample revealed that the Ag deposition did not degrade the critical current.

\begin{table*}[!t]
	\caption{Characteristics of b-CFD samples. Sample S4 is considered as the reference sample.}
	\begin{ruledtabular}
	\begin{tabular}{ccccc}
		   ~Sample ID~     & ~Top Ag thickness~      & ~Bottom Ag thickness~      & ~~~Ratio~~~   & ~Total Ag~thickness \vspace{0ex} \\
	       & ($t_{\text{top}}$, $\mu$m)   & ($t_{\text{bot}}$, $\mu$m)  & ($t_{\text{bot}}/t_{\text{top}}$)   & ($t_\text{tot}$, $\mu$m) \vspace{1ex} \\
	       \hline \\
	       S1\phantom{ (ref)}      & 0.02     & $\approx 1.98$     & 99            & 2.0   \vspace{0ex} \\      
	       S2\phantom{ (ref)}      & 0.1      & 1.9                & 19            & 2.0   \vspace{0ex} \\      
	       S3\phantom{ (ref)}      & 0.3      & 1.7                & 5.7           & 2.0   \vspace{0ex} \\      
	       S4 (ref)  			   & 2.0      & 0.0				   & 0.0		   & 2.0   \vspace{0ex} \\  
	\end{tabular}
	\end{ruledtabular}
	\label{tab:bCFD_samples}
\end{table*}

\subsection{Measurement of Normal Zone Propagation Velocity}
The NZPV has been measured using the same setup as described in~\cite{Lacroix2013a,Lacroix2014}. A home-made pulsed current source was used to generate square pulses ($\Delta t=10-20$ ms, $I=55-110$ A). A small NdFeB permanent magnet was used to generate a normal zone by lowering locally  the critical current $I_c$. The NZPV was measured using voltage taps in contact with the surface of the tape placed every $\approx2.5$~mm. By measuring the voltage drop generated by the normal zone over time at different locations along the tape, the velocity of the normal zone propagation could be obtained~\cite{Lacroix2013a}.
During all these NZPV measurements, the samples were placed in a liquid nitrogen bath.

\subsection{Video recording of the dynamic of the quench}
Furthermore, an optical in situ study of the quench propagation in liquid nitrogen was also realized, similarly to previous studies reported in the literature~\cite{Kraemer2003, Nam2006, Nguyen2009, Nguyen2010}. The technique consists of quenching locally an HTS tape immersed in liquid nitrogen in order to visualize the propagation of nitrogen gas bubbles. These bubbles, which are generated due to the heat generation at the normal zone location, induce light refraction and multiple reflections at liquid-gas interfaces. This modulates the light intensity received by the detector and create a contrast between the NZ and the superconducting part of the tape. During the experiment, a pulsed current is injected in the sample and a normal zone is generated by a permanent magnet placed behind the sample. A video is recorded using a MEMRECAM HX high-speed camera (CMOS sensor) from nac Image Technology and an AF Micro-NIKKOR (60 mm f/2.8D) lens from Nikon. The high-speed camera is placed behind the double-pane window of a cryostat to separate it from the liquid nitrogen bath. Special care has been taken to prevent bubbles from obstructing the field of view of the camera. 

\section{Electro-thermal model}
A 3-D finite element electro-thermal model has been used to simulate the quench behaviour of the samples. Details about the model can be found in~\cite{Lacroix2014a,Lacroix2017}. The numerical calculations have been realized with the Joule heating module of the COMSOL 4.3b software program, which solves simultaneously the heat equation and the current continuity equation. The coupling of the two equations is ensured through the Joule losses and through the temperature dependence of the electrical parameters. The $E-J$ characteristic of (RE)BaCuO was modelled using an empirical temperature dependent power law $E(J,T)=E_0\left(J/J_c(T)\right)^{n(T)}$ in the flux creep and flux flow regimes~\cite{Brandt1996}, where $E_0=1$ $\mu$V/cm (electric field criterion at 77~K). The transition from the superconducting state to the normal state was modelled as a normal state path in parallel with the superconducting one, such as two resistances in parallel in an electrical circuit. A non-linear conductivity is derived from the power law
model by rewriting it in a form compatible with the constitutive equation $J=\sigma E$, which gives

\begin{equation}
        \label{sigma}
        \sigma_{sc}(T)= \frac{J_c(T)}{E_0}\left(\frac{\left\|\textbf{E}\right\|}{E_0}\right)^{\frac{1-n(T)}{n(T)}} \,,
\end{equation}
where the temperature dependence of the power-law index $n(T)$ has been linearly extrapolated from~\cite{Polat2011a}, i.e.
\begin{equation}
        \label{n}
        n(T)= 
        \begin{cases}
                 (n_0-10)\left(\frac{T_c-T}{T_c-T_0}\right)+10 & \text{for } T<T_c \,, \\
                 10  & \text{for } T\geq T_c \,.
        \end{cases}
\end{equation}
The temperature dependence of $J_c(T)$ is considered linear with temperature, i.e.
\begin{equation}
        \label{Jc}
        J_c(T)= 
        \begin{cases}
                        J_{c0}\left(\frac{T_c-T}{T_c-T_0}\right) & \text{for } T<T_c \,, \\
                        0  & \text{for } T\geq T_c \,.
        \end{cases}
\end{equation}
The parameters $J_{c0}$ and $n_0$ have been extracted from experimental $I-V$ curves obtained for each sample. The values obtained for $J_{c0}$ and $n_0$ are respectively in the range of $2.2-2.6 \times 10^{10}$ A/m$^2$ and  $18.6 - 21.4$ at 77~K. Simulations were carried out for two values of interfacial resistance between Ag and (RE)BaCuO, i.e. 0.2 $\mu\Omega$.cm$^2$ and 1 $\mu\Omega$.cm$^2$. In the simulations, the normal zone has been initiated by applying a local heat pulse at one extremity of the tape. Virtual probes were placed every 1~mm in order to monitor the time evolution of the voltage and the temperature, from which the NZPV can be calculated.

\section{Results}
\subsection{Experiments}
\begin{figure*}[t!]
        \includegraphics[width=13cm]{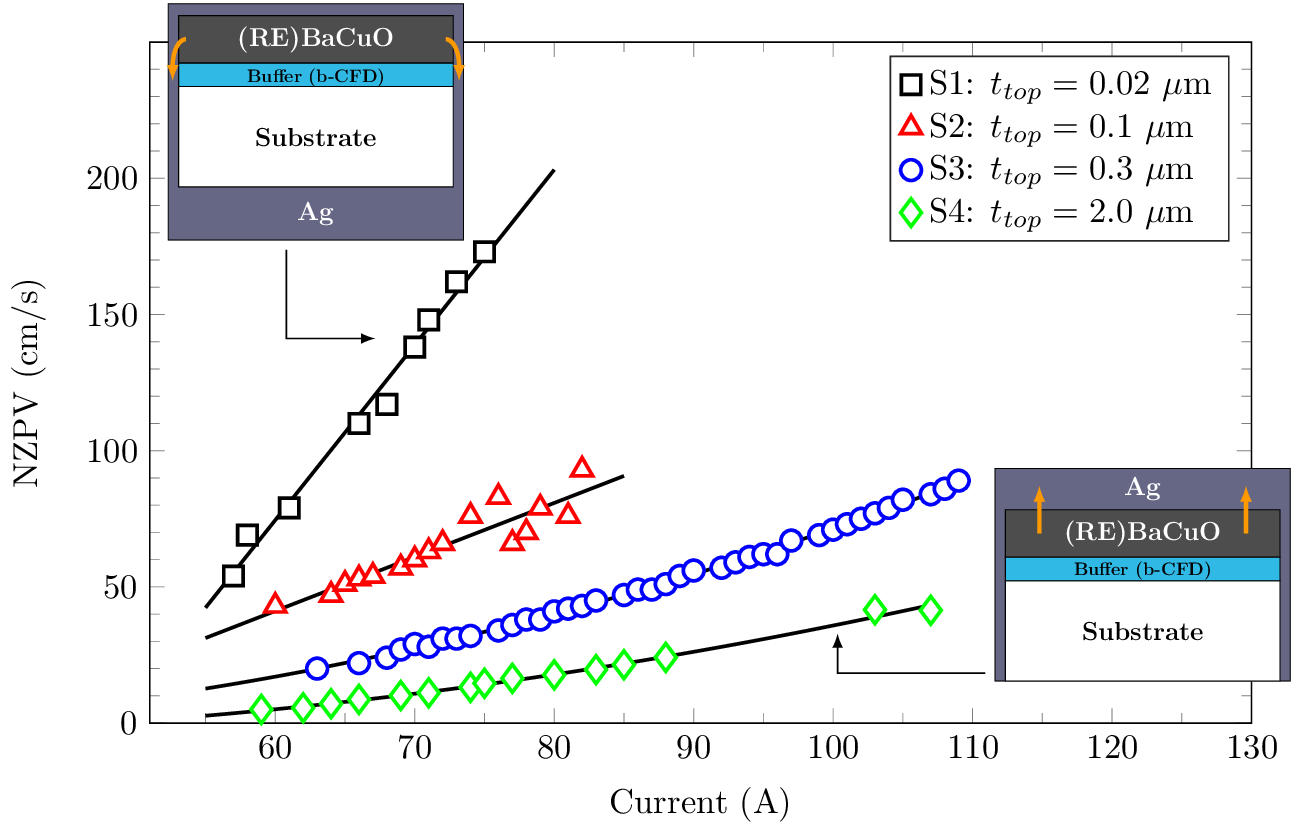}
        \caption{Normal zone propagation velocity as a function of the transport current for different stabilizer thicknesses on the superconductor side. Schematics of the cross-section views of the two extreme cases are shown in the insets. The case with $t_{\text{top}}=2~\mu$m of Ag (green diamonds) is comparable to commercial tapes (no CFD effect) and exhibits the lowest NZPV values. The case with $t_{\text{top}}=0.02~\mu$m of Ag (black squares) shows the highest NZPV values measured. Solid lines are quadratic functions fitted to the experimental data to illustrate the trend.
                        }
        \label{fig:2}
\end{figure*}

The NZPV has been measured experimentally for all samples described in table~\ref{tab:bCFD_samples}. Fig.~\ref{fig:2} presents the NZPV values as a function of the transport current for all four samples. First, we observe that the NZPV increases approximately linearly with the current, as expected by the adiabatic model~\cite{Dresner2002}. Indeed, S1 and S2 are well fitted by a linear function, while S3 and S4 are better fitted with a quadratic function,
although the curvature is very slight.

Furthermore, we observe that, for a given transport current, the NZPV is higher for samples having a thinner layer of Ag on the (RE)BaCuO side (top side) varying from 14 cm/s when $t_{\text{top}}=2~\mu$m (green diamonds) to 171 cm/s (black squares) when $t_{\text{top}}=0.02~\mu$m, for $I=75$~A.

\begin{figure}[h!]
        \includegraphics[width=8cm]{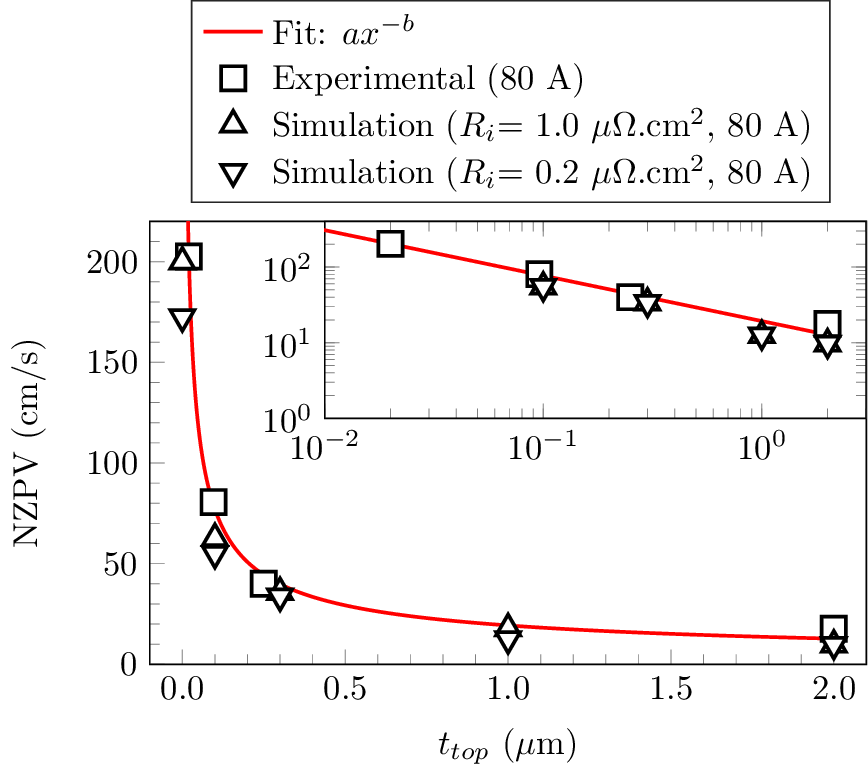}
        \caption{Plot of measured and simulated NZPV values as a function of the top Ag stabilizer thickness ($t_{\text{top}}$), for a fixed transport current of 80~A in all cases. The solid curve shows a fit of the form $y=ax^{-b}$, with $a=23.0918$ and $b=0.5333$. Inset: same plot in log-log scale.}
       \label{fig:3}
\end{figure}

Fig.~\ref{fig:3} present the NZPV values (squares) obtained experimentally for an applied current of 80~A. We observe that the NZPV increases drastically for small values of $t_{\text{top}}$, while it remains almost constant when $t_{\text{top}}$ is larger than $1~\mu$m. In the inset, the same results are presented using a logarithmic scale for the $x$ and  $y$ axes. The experimental data points can be well fitted with a function of the form $\text{NZPV}=a\cdot\left(t_{\text{top}}\right)^{-b}$, with $a=23.0918$ and $b=0.5333$.

Fig.~\ref{fig:4} presents snapshots of the nucleation of a normal zone induced by a permanent magnet of 3 mm in diameter placed behind the sample. The generation of the bubbles was recorded in the cases of sample S2 ($t_{\text{top}}=0.1$~$\mu$m) and sample S4 ($t_{\text{top}}=2.0$~$\mu$m). As observed in Fig.~\ref{fig:4}, the normal zone nucleates at the location of the magnet (vertical black line in the middle of the picture) and propagates in the two opposite directions, as indicated by the arrows in the close-up views. When comparing the two snapshots, we observe that the shapes of the normal zones are drastically different. A ``U-shaped'' profile is clearly visible in the case of sample S2 at both extremities of the normal zone, whereas a quasi-linear profile is observed in the case of sample S4. Those profiles stay roughly stationary while the normal zone is propagating (the propagation stops when the current pulse ends), then the tape slowly recovers (e.g. the normal zone shrinks until it disappears). 

\subsection{Comparison with numerical results}
In Fig.~\ref{fig:3}, we compare the NZPV values obtained experimentally with those obtained by numerical simulations for two values of contact resistance, i.e. $R_i = 0.2$ and $1~\mu\Omega$.cm$^2$. The value of $R_i$ can be ajusted by directly varying the conductivity of the contact resistance at the Ag/YBaCuO interface in the model. The smallest $R_i$ value corresponds to the experimental value measured previously on a commercial tape~\cite{Lacroix2013a}, whereas the highest one was used to quantify the effect of $R_i$ on the NZPV of tapes with the b-CFD architecture. We note that these values are typical for the case of Ag that has been sputtered or evaporated on (RE)BaCuO~\cite{Ekin2006}. We observe that our simulation results are in very good agreement with our experimental results. We also observe that the effect of $R_i$ on the NZPV is much more pronounced in the case of thin layers of Ag on the superconductor side, i.e. small values of $t_{\text{top}}$. Indeed, the numerical value of the NZPV obtained when $t_{\text{top}}=0~\mu$m and $R_i=1~\mu\Omega$.cm$^2$ is 200 cm/s, while for a lower interfacial resistance ($0.2~\mu\Omega$.cm$^2$), the simulation gives 173 cm/s. Considering that the NZPV value obtained experimentally in the case where $t_{\text{top}}\approx 20~$nm (sample S1) is 203 cm/s, it suggests that a value of 1~$\mu\Omega$.cm$^2$ for the interfacial resistance is more appropriate for the samples considered in the present work. 

We note that the difference between the NZPV values obtained with different interfacial resistances quickly becomes negligible as $t_{\text{top}}$ increases. When we look more closely at the experimental values vs.~the values obtained by numerical simulations, we notice that the simulations give in general slightly lower NZPV values. This difference could possibly be explained as follows. Firstly, the non-uniformity of the Ag layers deposited by sputtering, which is known to induce thickness variations in the range of $\pm \: 50$~nm (according to Dektak thickness profile measurements) is not taken into account in the numerical model. Secondly, the Ag thickness on the edges of the tape is not accurate. In our electro-thermal model, we assumed uniform Ag coating on both faces of the tapes, and a thickness of 1 $\mu$m for the Ag on the edges of the tape. The latter number was inferred from the fact that the total thickness of Ag deposited on each sample was $2~\mu$m, but each edge of the sample has been exposed only for one half of the total sputtering time in the chamber (because of the tilted position of the sample in the chamber, used to ensure a good deposition on the edges of the tape). These two factors can likely induce an error in the calculation of the local Joule losses and, therefore, in the calculation of the temperature and NZPV.

\begin{figure*}[]
	  \includegraphics[width=15cm]{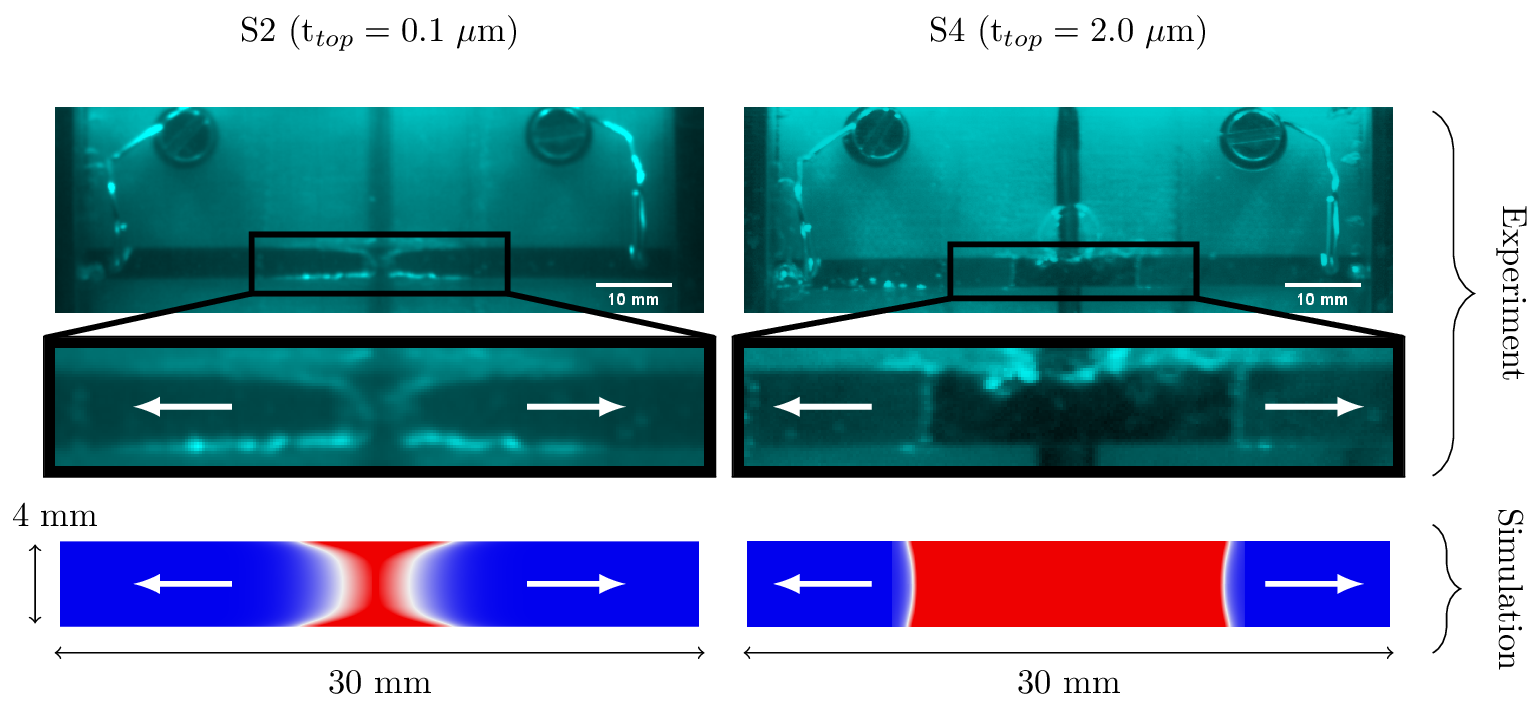}
      \caption{Visualization of the normal zone propagation obtained experimetally (snapshots of a video recorded with a digital high-speed CMOS camera), for two samples: S2 on the left, and S4 on the right. The snapshots show the sample (black horizontal strip) as well as the two voltage taps, at the ends of the sample, held by two screws. The vertical black line in the middle indicate the position of the permanent magnet behind the sample. Below each snapshot, we show a close-up view of the normal zone indicated by the generation of bubbles, which induces a change in the image contrast. The arrows indicate the direction of the normal zone propagation. The current waveform injected in the sample was a square pulse. The snapshot of S2 and S4 have been taken respectively 16~ms after the current had reached 60~A (recorded at 4000 frames/s), and 40~ms after the current had reached 70~A (recorded at 5000 frames/s). The distribution of temperature obtained by simulation is shown for comparison under the close-up views for both samples (the simulations were performed with the same parameters as the experimental ones). The normal zone is indicated in red ($T>T_c$) and the superconducting state is in blue ($T=77$~K). The thin white strips indicate the middle temperature between $T_c=90$~K and $T=77$~K, which is about 83.5 K.}
      \label{fig:4}
\end{figure*}

The simulated surface temperature distributions of the Ag layer (top view of the 3-D model) for each sample are shown in Fig.~\ref{fig:4} below the images taken from the video recording.The parameters used for the simulations are the same as the ones used in Fig.~\ref{fig:3} with $R_i=0.2~\mu\Omega.\text{cm}^2$. The simulations clearly exhibit the characteristic ``U-shaped'' profile between the normal zone ($T>T_c$ in red) and the superconducting state ($T=77$~K in blue) observed in the S2 snapshot. The creation of a U-shaped profile is easy to explain considering that, when a normal zone nucleates, the current flowing into the superconductor transfers into the stabilizer by the edges of the tape, as shown in Fig.~\ref{fig:1}c), which generates a very high instantaneous power density (see~\cite{Lacroix2017} for more details). The amount of heat that is generated in the edges of the tape during the quench forces the normal zone to initally nucleate in the edges of the tape. Also, we note that the simulations reproduce well the straight superconductor/normal boundary profile observed in the S4 snapshot.

\section{Discussion}
In a previous study, Levin et al.~\cite{Levin2007b} were able to find the form of the voltage distribution into the stabilizer in the case of an HTS tape surrounded by a stabilizer, similar to the b-CFD architecture presented in this paper. Their results showed that Joule losses are produced at the edges of the tape when the current transfers from the superconductor layer to the bottom stabilizer layer, similarly to our findings. 

Furthermore, as Levin et al. discussed in their paper, in the case of a large normal zone, the current flowing in the stabilizer fills equally both stabilizer layers (top and bottom) if the thickness of each side is the same, which is quite obvious from basic circuit theory. This means that we can model the two stabilizer layers as two resistances in parallel. Thus, in the case of the b-CFD architecture, since the bottom stabilizer layer is much thicker than the top stabilizer layer, most of the current flows in the bottom stabilizer layer.

\begin{figure}[h!]
      \includegraphics[width=8.5cm]{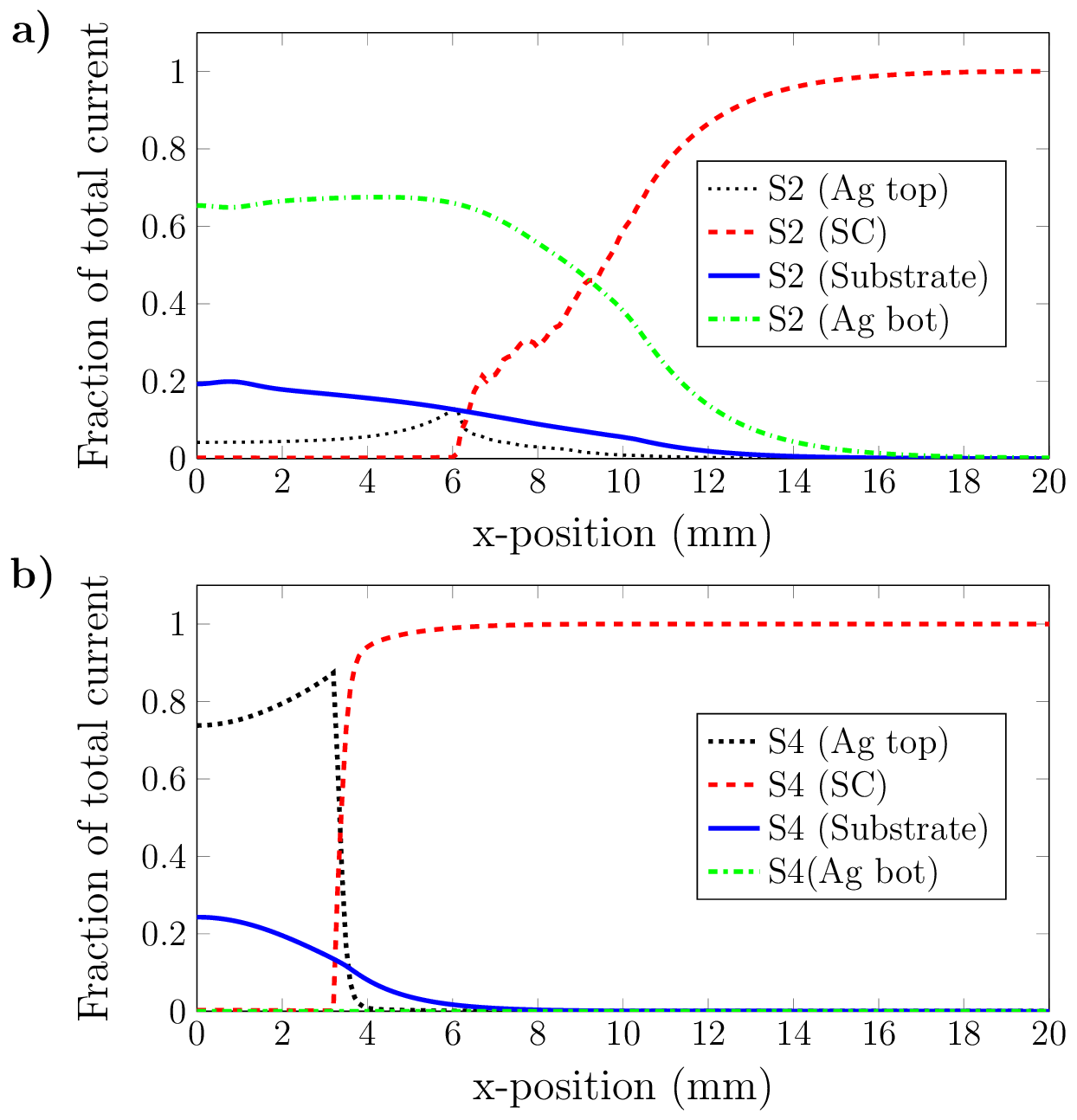}
      \caption{Total current flowing in the different layers of the HTS tape architecture, plotted along the tape length, for samples a) S2 and b) S4. Results were obtained by numerical simulations for tape architectures representative of samples S2 and S4, for $I=80$~A and $R_i=0.2~\mu\Omega.\text{cm}^2$. The current in each layer was calculated as the integral of the current density $J$ in the cross-section of the layer. In all cases, the normal zone was initiated at $x=0$. The data shown above were taken at a) $t=15$~ms and b) $t=30$~ms.}
      \label{fig:5}
\end{figure}

Fig.~\ref{fig:5} presents the current sharing between the different layers along the length of the tape (in the current direction) for samples S2 and S4 obtained from numerical simulations. In both cases, the current is flowing in the superconductor until it reaches the normal zone, and then it transfers into the stabilizer (top and bottom layers) and into the substrate. In the case of sample S2, we see from Fig.~\ref{fig:5}a) that most of the current transfers from the superconductor to the bottom stabilizer layer, while in the case of sample S4 (Fig.~\ref{fig:5}b)), most of the current transfers from the superconductor to the top stabilizer layer, which confirms the validity of the model based on two resistances in parallel.Furthermore, we note that, in the case of sample S2, the current transfers from the superconductor to the metallic layers over a length of approximately 10 mm. However, in the case of sample S4, we observe that it takes less than 1 mm for the current to transfer from the superconductor to the metallic layers (Fig.~\ref{fig:5}b)). This clearly shows that the ``b-CFD'' architecture increases the current transfer length (CTL), which is the key to increase the NZPV~\cite{Levin2010,Lacroix2017}.

The quenching behaviour of the HTS tapes investigated in this work can thus be understood by comparing the paths taking by the current when transferring from the (RE)BaCuO layer to the Ag stabilizer for the two extreme cases, i.e. $t_{\text{top}} = 0.02~\mu$m and $t_{\text{top}}=2.0~\mu$m. In the case where all Ag is deposited on the (RE)BaCuO side ($t_{\text{top}}=2~\mu$m), the current goes directly into the top Ag layer without seeing resistance, as it circumvents the quenched region. This architecture is typical of commercial tapes and possesses the lowest NZPV. When $t_{\text{top}}=0.02~\mu$m, almost all the current is forced to transfer to the bottom stabilizer through the Ag bridges located on the edges of the tape, and therefore a significant electrical resistance is seen by the current (see~\cite{Lacroix2017} for more details). This higher resistance increases the current transfer length (CTL) and thus, the NZPV~\cite{Levin2010}.

\section{Conclusion}
In this paper, we introduced an original architecture for the metallic thermal stabilizer required in (RE)BaCuO high temperature superconductor (HTS) tapes. This architecture was called ``b-CFD'', standing for ``buffer-Current Flow Diverter'', since it consists of an implementation of the current flow diverter concept~\cite{Lacroix2014} by using the electrical insulation properties of the buffer layers, already present in existing HTS tapes. Similarly as in the Current Flow Diverter (CFD) concept, this architecture allows increasing substantially the normal zone propagation velocity of the tapes by concentrating the generation of the Joule losses at specific locations in the tape when the current transfers between the superconducting and stabilizer layers. However, as opposed to the classical CFD concept, the b-CFD architecture does not increase the interfacial resistance between the stabilizer and the superconductor. Furthermore, the b-CFD architecture allows increasing the NZPV even in the presence of a thick stabilizer, which is not the case for the classical CFD architecture~\cite{Lacroix2017}. This is particularly interesting for applications requiring a thick stabilizer such as superconducting electromagnets.

Although the concept is not well proved, there remains a challenge for finding a way to process such conductors in long lengths. Indeed, the need to ensure an electrical connection between the top and bottom stabilizers along hundreds of meters of tape, combined with the very low thickness ($\approx 20$ to 100~nm) required for the top Ag stabilizer (which tends to form aggregates under a heat treatment) requires further thoughts. Note however that the latter problem was not encountered with our samples from STI, because their process does not require any oxygen annealing of the HTS tapes after their fabrication.

\section{Acknowledgements}
The authors gratefully thank the support of Superconductor Technologies Inc. (STI), in particular Ken Pfeiffer and Jeong Huh, for the production of the custom-made samples that allowed realizing this experiment and proving the b-CFD architecture concept. This work was supported by research grants from NSERC (Canada) and FRQNT (Qu\'ebec).
\\ \\ \\ \\ \\ \\ \\ \\ \\ \\ \\ \\ \\

\bibliographystyle{apsrev4-1.bst}

\begin{thebibliography}{29}%
	\makeatletter
	\providecommand \@ifxundefined [1]{\@ifx{#1\undefined}}
	\providecommand \@ifnum [1]{\ifnum #1\expandafter \@firstoftwo \else \expandafter \@secondoftwo \fi}
	\providecommand \@ifx [1]{\ifx #1\expandafter \@firstoftwo \else \expandafter \@secondoftwo \fi}
	\providecommand \natexlab [1]{#1}%
	\providecommand \enquote  [1]{``#1''}%
	\providecommand \bibnamefont  [1]{#1}%
	\providecommand \bibfnamefont [1]{#1}%
	\providecommand \citenamefont [1]{#1}%
	\providecommand \href@noop [0]{\@secondoftwo}%
	\providecommand \href [0]{\begingroup \@sanitize@url \@href}%
	\providecommand \@href[1]{\@@startlink{#1}\@@href}%
	\providecommand \@@href[1]{\endgroup#1\@@endlink}%
	\providecommand \@sanitize@url [0]{\catcode `\\12\catcode `\$12\catcode
	  `\&12\catcode `\#12\catcode `\^12\catcode `\_12\catcode `\%12\relax}%
	\providecommand \@@startlink[1]{}%
	\providecommand \@@endlink[0]{}%
	\providecommand \url  [0]{\begingroup\@sanitize@url \@url }%
	\providecommand \@url [1]{\endgroup\@href {#1}{\urlprefix }}%
	\providecommand \urlprefix  [0]{URL }%
	\providecommand \Eprint [0]{\href }%
	\providecommand \doibase [0]{http://dx.doi.org/}%
	\providecommand \selectlanguage [0]{\@gobble}%
	\providecommand \bibinfo  [0]{\@secondoftwo}%
	\providecommand \bibfield  [0]{\@secondoftwo}%
	\providecommand \translation [1]{[#1]}%
	\providecommand \BibitemOpen [0]{}%
	\providecommand \bibitemStop [0]{}%
	\providecommand \bibitemNoStop [0]{.\EOS\space}%
	\providecommand \EOS [0]{\spacefactor3000\relax}%
	\providecommand \BibitemShut  [1]{\csname bibitem#1\endcsname}%
	\let\auto@bib@innerbib\@empty
	\bibitem [{\citenamefont {Gurevich}(2001)}]{Gurevich2001}%
	  \BibitemOpen
	  \bibfield  {author} {\bibinfo {author} {\bibfnamefont {A.}~\bibnamefont
	  {Gurevich}},\ }\href {\doibase 10.1063/1.1358361} {\bibfield  {journal}
	  {\bibinfo  {journal} {Applied Physics Letters}\ }\textbf {\bibinfo {volume}
	  {78}},\ \bibinfo {pages} {1891} (\bibinfo {year} {2001})}\BibitemShut
	  {NoStop}%
	\bibitem [{\citenamefont {Furtner}\ \emph {et~al.}(2004)\citenamefont
	  {Furtner}, \citenamefont {Nemetschek}, \citenamefont {Semerad}, \citenamefont
	  {Sigl},\ and\ \citenamefont {Prusseit}}]{Furtner2004}%
	  \BibitemOpen
	  \bibfield  {author} {\bibinfo {author} {\bibfnamefont {S.}~\bibnamefont
	  {Furtner}}, \bibinfo {author} {\bibfnamefont {R.}~\bibnamefont {Nemetschek}},
	  \bibinfo {author} {\bibfnamefont {R.}~\bibnamefont {Semerad}}, \bibinfo
	  {author} {\bibfnamefont {G.}~\bibnamefont {Sigl}}, \ and\ \bibinfo {author}
	  {\bibfnamefont {W.}~\bibnamefont {Prusseit}},\ }\href {\doibase
	  10.1088/0953-2048/17/5/037} {\bibfield  {journal} {\bibinfo  {journal}
	  {Superconductor Science and Technology}\ }\textbf {\bibinfo {volume} {17}},\
	  \bibinfo {pages} {S281} (\bibinfo {year} {2004})}\BibitemShut {NoStop}%
	\bibitem [{\citenamefont {Kudymow}\ \emph {et~al.}(2007)\citenamefont
	  {Kudymow}, \citenamefont {Noe}, \citenamefont {Schacherer}, \citenamefont
	  {Kinder},\ and\ \citenamefont {Prusseit}}]{Kudymow2007}%
	  \BibitemOpen
	  \bibfield  {author} {\bibinfo {author} {\bibfnamefont {A.}~\bibnamefont
	  {Kudymow}}, \bibinfo {author} {\bibfnamefont {M.}~\bibnamefont {Noe}},
	  \bibinfo {author} {\bibfnamefont {C.}~\bibnamefont {Schacherer}}, \bibinfo
	  {author} {\bibfnamefont {H.}~\bibnamefont {Kinder}}, \ and\ \bibinfo {author}
	  {\bibfnamefont {W.}~\bibnamefont {Prusseit}},\ }\href {\doibase
	  10.1109/TASC.2007.899578} {\bibfield  {journal} {\bibinfo  {journal} {IEEE
	  Transactions on Applied Superconductivity}\ }\textbf {\bibinfo {volume}
	  {17}},\ \bibinfo {pages} {3499} (\bibinfo {year} {2007})}\BibitemShut
	  {NoStop}%
	\bibitem [{\citenamefont {Colangelo}\ and\ \citenamefont
	  {Dutoit}(2012)}]{Colangelo2012a}%
	  \BibitemOpen
	  \bibfield  {author} {\bibinfo {author} {\bibfnamefont {D.}~\bibnamefont
	  {Colangelo}}\ and\ \bibinfo {author} {\bibfnamefont {B.}~\bibnamefont
	  {Dutoit}},\ }\href {\doibase 10.1088/0953-2048/25/9/095005} {\bibfield
	  {journal} {\bibinfo  {journal} {Superconductor Science and Technology}\
	  }\textbf {\bibinfo {volume} {25}},\ \bibinfo {pages} {095005} (\bibinfo
	  {year} {2012})}\BibitemShut {NoStop}%
	\bibitem [{\citenamefont {Lacroix}\ \emph {et~al.}(2013)\citenamefont
	  {Lacroix}, \citenamefont {Fournier-Lupien}, \citenamefont {Mcmeekin},\ and\
	  \citenamefont {Sirois}}]{Lacroix2013a}%
	  \BibitemOpen
	  \bibfield  {author} {\bibinfo {author} {\bibfnamefont {C.}~\bibnamefont
	  {Lacroix}}, \bibinfo {author} {\bibfnamefont {J.-H.}\ \bibnamefont
	  {Fournier-Lupien}}, \bibinfo {author} {\bibfnamefont {K.}~\bibnamefont
	  {Mcmeekin}}, \ and\ \bibinfo {author} {\bibfnamefont {F.}~\bibnamefont
	  {Sirois}},\ }\href@noop {} {\bibfield  {journal} {\bibinfo  {journal} {IEEE
	  Transactions on Applied Superconductivity}\ }\textbf {\bibinfo {volume}
	  {23}},\ \bibinfo {pages} {4701605} (\bibinfo {year} {2013})}\BibitemShut
	  {NoStop}%
	\bibitem [{\citenamefont {Lacroix}\ \emph {et~al.}(2014)\citenamefont
	  {Lacroix}, \citenamefont {Lapierre}, \citenamefont {Coulombe},\ and\
	  \citenamefont {Sirois}}]{Lacroix2014}%
	  \BibitemOpen
	  \bibfield  {author} {\bibinfo {author} {\bibfnamefont {C.}~\bibnamefont
	  {Lacroix}}, \bibinfo {author} {\bibfnamefont {Y.}~\bibnamefont {Lapierre}},
	  \bibinfo {author} {\bibfnamefont {J.}~\bibnamefont {Coulombe}}, \ and\
	  \bibinfo {author} {\bibfnamefont {F.}~\bibnamefont {Sirois}},\ }\href
	  {\doibase 10.1088/0953-2048/27/5/055013} {\bibfield  {journal} {\bibinfo
	  {journal} {Superconductor Science and Technology}\ }\textbf {\bibinfo
	  {volume} {27}},\ \bibinfo {pages} {055013} (\bibinfo {year}
	  {2014})}\BibitemShut {NoStop}%
	\bibitem [{\citenamefont {Lacroix}\ and\ \citenamefont
	  {Sirois}(2014)}]{Lacroix2014a}%
	  \BibitemOpen
	  \bibfield  {author} {\bibinfo {author} {\bibfnamefont {C.}~\bibnamefont
	  {Lacroix}}\ and\ \bibinfo {author} {\bibfnamefont {F.}~\bibnamefont
	  {Sirois}},\ }\href {\doibase 10.1088/0953-2048/27/3/035003} {\bibfield
	  {journal} {\bibinfo  {journal} {Superconductor Science and Technology}\
	  }\textbf {\bibinfo {volume} {27}},\ \bibinfo {pages} {035003} (\bibinfo
	  {year} {2014})}\BibitemShut {NoStop}%
	\bibitem [{\citenamefont {Levin}\ \emph {et~al.}(2010)\citenamefont {Levin},
	  \citenamefont {Jones}, \citenamefont {Novak},\ and\ \citenamefont
	  {Barnes}}]{Levin2010}%
	  \BibitemOpen
	  \bibfield  {author} {\bibinfo {author} {\bibfnamefont {G.~A.}\ \bibnamefont
	  {Levin}}, \bibinfo {author} {\bibfnamefont {W.~A.}\ \bibnamefont {Jones}},
	  \bibinfo {author} {\bibfnamefont {K.~A.}\ \bibnamefont {Novak}}, \ and\
	  \bibinfo {author} {\bibfnamefont {P.~N.}\ \bibnamefont {Barnes}},\
	  }\href@noop {} {\bibfield  {journal} {\bibinfo  {journal} {Superconductor
	  Science and Technology}\ }\textbf {\bibinfo {volume} {23}},\ \bibinfo {pages}
	  {014021} (\bibinfo {year} {2010})}\BibitemShut {NoStop}%
	\bibitem [{\citenamefont {Duckworth}(2001)}]{Duckworth2001}%
	  \BibitemOpen
	  \bibfield  {author} {\bibinfo {author} {\bibfnamefont {R.~C.}\ \bibnamefont
	  {Duckworth}},\ }\emph {\bibinfo {title} {{Contact resistance and normal zone
	  formation in coated yttrium barium copper oxide superconductors}}},\ \href
	  {http://www.asc.wisc.edu/pdf{\_}papers/theses/rcd01phd.pdf} {Ph.D. thesis},\
	  \bibinfo  {school} {University of Wisconsin-Madison} (\bibinfo {year}
	  {2001})\BibitemShut {NoStop}%
	\bibitem [{\citenamefont {Grabovickic}\ \emph {et~al.}(2003)\citenamefont
	  {Grabovickic}, \citenamefont {Lue}, \citenamefont {Gouge}, \citenamefont
	  {Demko},\ and\ \citenamefont {Duckworth}}]{Grabovickic2003}%
	  \BibitemOpen
	  \bibfield  {author} {\bibinfo {author} {\bibfnamefont {R.}~\bibnamefont
	  {Grabovickic}}, \bibinfo {author} {\bibfnamefont {J.~W.}\ \bibnamefont
	  {Lue}}, \bibinfo {author} {\bibfnamefont {M.~J.}\ \bibnamefont {Gouge}},
	  \bibinfo {author} {\bibfnamefont {J.~A.}\ \bibnamefont {Demko}}, \ and\
	  \bibinfo {author} {\bibfnamefont {R.~C.}\ \bibnamefont {Duckworth}},\ }\href
	  {\doibase 10.1109/TASC.2003.812874} {\bibfield  {journal} {\bibinfo
	  {journal} {IEEE Transactions on Applied Superconductivity}\ }\textbf
	  {\bibinfo {volume} {13}},\ \bibinfo {pages} {1726} (\bibinfo {year}
	  {2003})}\BibitemShut {NoStop}%
	\bibitem [{\citenamefont {Wang}\ \emph {et~al.}(2005)\citenamefont {Wang},
	  \citenamefont {Caruso}, \citenamefont {Breschi}, \citenamefont {Zhang},
	  \citenamefont {Trociewitz}, \citenamefont {Weijers},\ and\ \citenamefont
	  {Schwartz}}]{Wang2005a}%
	  \BibitemOpen
	  \bibfield  {author} {\bibinfo {author} {\bibfnamefont {X.}~\bibnamefont
	  {Wang}}, \bibinfo {author} {\bibfnamefont {A.~R.}\ \bibnamefont {Caruso}},
	  \bibinfo {author} {\bibfnamefont {M.}~\bibnamefont {Breschi}}, \bibinfo
	  {author} {\bibfnamefont {G.}~\bibnamefont {Zhang}}, \bibinfo {author}
	  {\bibfnamefont {U.~P.}\ \bibnamefont {Trociewitz}}, \bibinfo {author}
	  {\bibfnamefont {H.~W.}\ \bibnamefont {Weijers}}, \ and\ \bibinfo {author}
	  {\bibfnamefont {J.}~\bibnamefont {Schwartz}},\ }\href {\doibase
	  10.1109/TASC.2005.847661} {\bibfield  {journal} {\bibinfo  {journal} {IEEE
	  Transactions on Applied Superconductivity}\ }\textbf {\bibinfo {volume}
	  {15}},\ \bibinfo {pages} {2586} (\bibinfo {year} {2005})}\BibitemShut
	  {NoStop}%
	\bibitem [{\citenamefont {Armenio}\ \emph {et~al.}(2008)\citenamefont
	  {Armenio}, \citenamefont {Augieri}, \citenamefont {Celentano}, \citenamefont
	  {Galluzzi}, \citenamefont {Mancini}, \citenamefont {Rufoloni}, \citenamefont
	  {Vannozzi}, \citenamefont {Gambardella}, \citenamefont {Saggese},
	  \citenamefont {Sessa},\ and\ \citenamefont {Pace}}]{Armenio2008}%
	  \BibitemOpen
	  \bibfield  {author} {\bibinfo {author} {\bibfnamefont {A.~A.}\ \bibnamefont
	  {Armenio}}, \bibinfo {author} {\bibfnamefont {A.}~\bibnamefont {Augieri}},
	  \bibinfo {author} {\bibfnamefont {G.}~\bibnamefont {Celentano}}, \bibinfo
	  {author} {\bibfnamefont {V.}~\bibnamefont {Galluzzi}}, \bibinfo {author}
	  {\bibfnamefont {A.}~\bibnamefont {Mancini}}, \bibinfo {author} {\bibfnamefont
	  {A.}~\bibnamefont {Rufoloni}}, \bibinfo {author} {\bibfnamefont
	  {A.}~\bibnamefont {Vannozzi}}, \bibinfo {author} {\bibfnamefont
	  {U.}~\bibnamefont {Gambardella}}, \bibinfo {author} {\bibfnamefont
	  {A.}~\bibnamefont {Saggese}}, \bibinfo {author} {\bibfnamefont
	  {P.}~\bibnamefont {Sessa}}, \ and\ \bibinfo {author} {\bibfnamefont
	  {S.}~\bibnamefont {Pace}},\ }\href {\doibase 10.1109/TASC.2008.920835}
	  {\bibfield  {journal} {\bibinfo  {journal} {IEEE Transactions on Applied
	  Superconductivity}\ }\textbf {\bibinfo {volume} {18}},\ \bibinfo {pages}
	  {1293} (\bibinfo {year} {2008})}\BibitemShut {NoStop}%
	\bibitem [{\citenamefont {Wang}\ \emph {et~al.}(2009)\citenamefont {Wang},
	  \citenamefont {Trociewitz},\ and\ \citenamefont {Schwartz}}]{Wang2009}%
	  \BibitemOpen
	  \bibfield  {author} {\bibinfo {author} {\bibfnamefont {X.}~\bibnamefont
	  {Wang}}, \bibinfo {author} {\bibfnamefont {U.~P.}\ \bibnamefont
	  {Trociewitz}}, \ and\ \bibinfo {author} {\bibfnamefont {J.}~\bibnamefont
	  {Schwartz}},\ }\href {\doibase 10.1088/0953-2048/22/8/085005} {\bibfield
	  {journal} {\bibinfo  {journal} {Supercond. Sci. Technol}\ }\textbf {\bibinfo
	  {volume} {22}},\ \bibinfo {pages} {085005} (\bibinfo {year}
	  {2009})}\BibitemShut {NoStop}%
	\bibitem [{\citenamefont {Park}\ \emph {et~al.}(2010)\citenamefont {Park},
	  \citenamefont {Kim}, \citenamefont {Park}, \citenamefont {Yu}, \citenamefont
	  {Eom}, \citenamefont {Bae}, \citenamefont {Kim}, \citenamefont {Sim},\ and\
	  \citenamefont {Sohn}}]{Park2010}%
	  \BibitemOpen
	  \bibfield  {author} {\bibinfo {author} {\bibfnamefont {H.-Y.}\ \bibnamefont
	  {Park}}, \bibinfo {author} {\bibfnamefont {A.-R.}\ \bibnamefont {Kim}},
	  \bibinfo {author} {\bibfnamefont {M.}~\bibnamefont {Park}}, \bibinfo {author}
	  {\bibfnamefont {I.-K.}\ \bibnamefont {Yu}}, \bibinfo {author} {\bibfnamefont
	  {B.-Y.}\ \bibnamefont {Eom}}, \bibinfo {author} {\bibfnamefont {J.-H.}\
	  \bibnamefont {Bae}}, \bibinfo {author} {\bibfnamefont {S.-H.}\ \bibnamefont
	  {Kim}}, \bibinfo {author} {\bibfnamefont {K.}~\bibnamefont {Sim}}, \ and\
	  \bibinfo {author} {\bibfnamefont {M.-H.}\ \bibnamefont {Sohn}},\ }\href@noop
	  {} {\bibfield  {journal} {\bibinfo  {journal} {IEEE Transactions on Applied
	  Superconductivity}\ }\textbf {\bibinfo {volume} {20}},\ \bibinfo {pages}
	  {2122} (\bibinfo {year} {2010})}\BibitemShut {NoStop}%
	\bibitem [{\citenamefont {Lu}\ \emph {et~al.}(2013)\citenamefont {Lu},
	  \citenamefont {Fang}, \citenamefont {Li}, \citenamefont {Wu},\ and\
	  \citenamefont {Guo}}]{Lu2013}%
	  \BibitemOpen
	  \bibfield  {author} {\bibinfo {author} {\bibfnamefont {W.~J.}\ \bibnamefont
	  {Lu}}, \bibinfo {author} {\bibfnamefont {J.}~\bibnamefont {Fang}}, \bibinfo
	  {author} {\bibfnamefont {D.}~\bibnamefont {Li}}, \bibinfo {author}
	  {\bibfnamefont {C.~Y.}\ \bibnamefont {Wu}}, \ and\ \bibinfo {author}
	  {\bibfnamefont {L.~J.}\ \bibnamefont {Guo}},\ }\href {\doibase
	  10.1016/j.physc.2012.03.062} {\bibfield  {journal} {\bibinfo  {journal}
	  {Physica C: Superconductivity}\ }\textbf {\bibinfo {volume} {484}},\ \bibinfo
	  {pages} {153} (\bibinfo {year} {2013})}\BibitemShut {NoStop}%
	\bibitem [{\citenamefont {Maeda}\ and\ \citenamefont
	  {Yanagisawa}(2014)}]{Maeda2014}%
	  \BibitemOpen
	  \bibfield  {author} {\bibinfo {author} {\bibfnamefont {H.}~\bibnamefont
	  {Maeda}}\ and\ \bibinfo {author} {\bibfnamefont {Y.}~\bibnamefont
	  {Yanagisawa}},\ }\href {\doibase 10.1109/TASC.2013.2287707} {\bibfield
	  {journal} {\bibinfo  {journal} {IEEE Transactions on Applied
	  Superconductivity}\ }\textbf {\bibinfo {volume} {24}},\ \bibinfo {pages} {1}
	  (\bibinfo {year} {2014})}\BibitemShut {NoStop}%
	\bibitem [{\citenamefont {Chan}\ \emph {et~al.}(2009)\citenamefont {Chan},
	  \citenamefont {Masson}, \citenamefont {Luongo},\ and\ \citenamefont
	  {Schwartz}}]{Chan2009}%
	  \BibitemOpen
	  \bibfield  {author} {\bibinfo {author} {\bibfnamefont {W.-K.}\ \bibnamefont
	  {Chan}}, \bibinfo {author} {\bibfnamefont {P.~J.}\ \bibnamefont {Masson}},
	  \bibinfo {author} {\bibfnamefont {C.~A.}\ \bibnamefont {Luongo}}, \ and\
	  \bibinfo {author} {\bibfnamefont {J.}~\bibnamefont {Schwartz}},\ }\href
	  {\doibase 10.1109/TASC.2009.2018514} {\bibfield  {journal} {\bibinfo
	  {journal} {IEEE Transactions on Applied Superconductivity}\ }\textbf
	  {\bibinfo {volume} {19}},\ \bibinfo {pages} {2490} (\bibinfo {year}
	  {2009})}\BibitemShut {NoStop}%
	\bibitem [{\citenamefont {Lacroix}\ \emph {et~al.}(2017)\citenamefont
	  {Lacroix}, \citenamefont {Sirois},\ and\ \citenamefont {{Fournier
	  Lupien}}}]{Lacroix2017}%
	  \BibitemOpen
	  \bibfield  {author} {\bibinfo {author} {\bibfnamefont {C.}~\bibnamefont
	  {Lacroix}}, \bibinfo {author} {\bibfnamefont {F.}~\bibnamefont {Sirois}}, \
	  and\ \bibinfo {author} {\bibfnamefont {J.-H.}\ \bibnamefont {{Fournier
	  Lupien}}},\ }\href@noop {} {\bibfield  {journal} {\bibinfo  {journal}
	  {Superconductor Science and Technology}\ } \textbf{\bibinfo {volume} {19}}, \bibinfo {pages} {064004} (\bibinfo {year}
	  {2017})}\BibitemShut {NoStop}%
	\bibitem [{Note1()}]{Note1}%
	  \BibitemOpen
	  \bibinfo {note} {Superconductor Technologies Inc., (http://www.suptech.com/),
	  Accessed November 11, 2016}\BibitemShut {NoStop}%
	\bibitem [{\citenamefont {Matias}\ \emph {et~al.}(2009)\citenamefont {Matias},
	  \citenamefont {Rowley}, \citenamefont {Coulter}, \citenamefont {Maiorov},
	  \citenamefont {Holesinger}, \citenamefont {Yung}, \citenamefont {Glyantsev},\
	  and\ \citenamefont {Moeckly}}]{Matias2009a}%
	  \BibitemOpen
	  \bibfield  {author} {\bibinfo {author} {\bibfnamefont {V.}~\bibnamefont
	  {Matias}}, \bibinfo {author} {\bibfnamefont {E.~J.}\ \bibnamefont {Rowley}},
	  \bibinfo {author} {\bibfnamefont {Y.}~\bibnamefont {Coulter}}, \bibinfo
	  {author} {\bibfnamefont {B.}~\bibnamefont {Maiorov}}, \bibinfo {author}
	  {\bibfnamefont {T.}~\bibnamefont {Holesinger}}, \bibinfo {author}
	  {\bibfnamefont {C.}~\bibnamefont {Yung}}, \bibinfo {author} {\bibfnamefont
	  {V.}~\bibnamefont {Glyantsev}}, \ and\ \bibinfo {author} {\bibfnamefont
	  {B.}~\bibnamefont {Moeckly}},\ }\href {\doibase
	  10.1088/0953-2048/23/1/014018} {\bibfield  {journal} {\bibinfo  {journal}
	  {Superconductor Science and Technology}\ }\textbf {\bibinfo {volume} {23}},\
	  \bibinfo {pages} {014018} (\bibinfo {year} {2009})}\BibitemShut {NoStop}%
	\bibitem [{\citenamefont {Ekin}(2006)}]{Ekin2006}%
	  \BibitemOpen
	  \bibfield  {author} {\bibinfo {author} {\bibfnamefont {J.~W.}\ \bibnamefont
	  {Ekin}},\ }\href {\doibase 10.1063/1.2743130} {\emph {\bibinfo {title}
	  {{Experimental Technicques for Low-Temperature Measurements}}}}\ (\bibinfo
	  {publisher} {Oxford university Press},\ \bibinfo {address} {Oxford},\
	  \bibinfo {year} {2006})\ Chap.\ \bibinfo {chapter} {8.2}, p.\ \bibinfo
	  {pages} {673}\BibitemShut {NoStop}%
	\bibitem [{\citenamefont {Kraemer}\ \emph {et~al.}(2003)\citenamefont
	  {Kraemer}, \citenamefont {Schmidt}, \citenamefont {Utz},\ and\ \citenamefont
	  {Neumueller}}]{Kraemer2003}%
	  \BibitemOpen
	  \bibfield  {author} {\bibinfo {author} {\bibfnamefont {H.~P.}\ \bibnamefont
	  {Kraemer}}, \bibinfo {author} {\bibfnamefont {W.}~\bibnamefont {Schmidt}},
	  \bibinfo {author} {\bibfnamefont {B.}~\bibnamefont {Utz}}, \ and\ \bibinfo
	  {author} {\bibfnamefont {H.~W.}\ \bibnamefont {Neumueller}},\ }\href
	  {\doibase 10.1109/TASC.2003.812980} {\bibfield  {journal} {\bibinfo
	  {journal} {IEEE Transactions on Applied Superconductivity}\ }\textbf
	  {\bibinfo {volume} {13}},\ \bibinfo {pages} {2044} (\bibinfo {year}
	  {2003})}\BibitemShut {NoStop}%
	\bibitem [{\citenamefont {Nam}\ \emph {et~al.}(2006)\citenamefont {Nam},
	  \citenamefont {Kang}, \citenamefont {Lee}, \citenamefont {Ko},\ and\
	  \citenamefont {Seok}}]{Nam2006}%
	  \BibitemOpen
	  \bibfield  {author} {\bibinfo {author} {\bibfnamefont {K.}~\bibnamefont
	  {Nam}}, \bibinfo {author} {\bibfnamefont {H.}~\bibnamefont {Kang}}, \bibinfo
	  {author} {\bibfnamefont {C.}~\bibnamefont {Lee}}, \bibinfo {author}
	  {\bibfnamefont {T.~K.}\ \bibnamefont {Ko}}, \ and\ \bibinfo {author}
	  {\bibfnamefont {B.~Y.}\ \bibnamefont {Seok}},\ }\href {\doibase
	  10.1109/TASC.2005.869669} {\bibfield  {journal} {\bibinfo  {journal} {IEEE
	  Transactions on Applied Superconductivity}\ }\textbf {\bibinfo {volume}
	  {16}},\ \bibinfo {pages} {727} (\bibinfo {year} {2006})}\BibitemShut
	  {NoStop}%
	\bibitem [{\citenamefont {Nguyen}\ and\ \citenamefont
	  {Tixador}(2009)}]{Nguyen2009}%
	  \BibitemOpen
	  \bibfield  {author} {\bibinfo {author} {\bibfnamefont {N.~T.}\ \bibnamefont
	  {Nguyen}}\ and\ \bibinfo {author} {\bibfnamefont {P.}~\bibnamefont
	  {Tixador}},\ }\href {\doibase 10.1088/0953-2048/23/2/025008} {\bibfield
	  {journal} {\bibinfo  {journal} {Superconductor Science and Technology}\
	  }\textbf {\bibinfo {volume} {23}},\ \bibinfo {pages} {25008} (\bibinfo {year}
	  {2009})}\BibitemShut {NoStop}%
	\bibitem [{\citenamefont {Nguyen}\ \emph {et~al.}(2010)\citenamefont {Nguyen},
	  \citenamefont {Barnier},\ and\ \citenamefont {Tixador}}]{Nguyen2010}%
	  \BibitemOpen
	  \bibfield  {author} {\bibinfo {author} {\bibfnamefont {N.~T.}\ \bibnamefont
	  {Nguyen}}, \bibinfo {author} {\bibfnamefont {C.}~\bibnamefont {Barnier}}, \
	  and\ \bibinfo {author} {\bibfnamefont {P.}~\bibnamefont {Tixador}},\ }\href
	  {\doibase 10.1088/1742-6596/234/3/032058} {\bibfield  {journal} {\bibinfo
	  {journal} {Journal of Physics: Conference Series}\ }\textbf {\bibinfo
	  {volume} {234}},\ \bibinfo {pages} {032058} (\bibinfo {year}
	  {2010})}\BibitemShut {NoStop}%
	\bibitem [{\citenamefont {Brandt}(1996)}]{Brandt1996}%
	  \BibitemOpen
	  \bibfield  {author} {\bibinfo {author} {\bibfnamefont {E.~H.}\ \bibnamefont
	  {Brandt}},\ }\href {\doibase 10.1103/PhysRevB.54.4246} {\bibfield  {journal}
	  {\bibinfo  {journal} {Physical Review B}\ }\textbf {\bibinfo {volume} {54}},\
	  \bibinfo {pages} {4246} (\bibinfo {year} {1996})}\BibitemShut {NoStop}%
	\bibitem [{\citenamefont {Polat}\ \emph {et~al.}(2011)\citenamefont {Polat},
	  \citenamefont {Sinclair}, \citenamefont {Zuev}, \citenamefont {Thompson},
	  \citenamefont {Christen}, \citenamefont {Cook}, \citenamefont {Kumar},
	  \citenamefont {Chen},\ and\ \citenamefont {Selvamanickam}}]{Polat2011a}%
	  \BibitemOpen
	  \bibfield  {author} {\bibinfo {author} {\bibfnamefont {O.}~\bibnamefont
	  {Polat}}, \bibinfo {author} {\bibfnamefont {J.~W.}\ \bibnamefont {Sinclair}},
	  \bibinfo {author} {\bibfnamefont {Y.~L.}\ \bibnamefont {Zuev}}, \bibinfo
	  {author} {\bibfnamefont {J.~R.}\ \bibnamefont {Thompson}}, \bibinfo {author}
	  {\bibfnamefont {D.~K.}\ \bibnamefont {Christen}}, \bibinfo {author}
	  {\bibfnamefont {S.~W.}\ \bibnamefont {Cook}}, \bibinfo {author}
	  {\bibfnamefont {D.}~\bibnamefont {Kumar}}, \bibinfo {author} {\bibfnamefont
	  {Y.}~\bibnamefont {Chen}}, \ and\ \bibinfo {author} {\bibfnamefont
	  {V.}~\bibnamefont {Selvamanickam}},\ }\href {\doibase
	  10.1103/PhysRevB.84.024519} {\bibfield  {journal} {\bibinfo  {journal}
	  {Physical Review B - Condensed Matter and Materials Physics}\ }\textbf
	  {\bibinfo {volume} {84}},\ \bibinfo {pages} {024519} (\bibinfo {year}
	  {2011})}\BibitemShut {NoStop}%
	\bibitem [{\citenamefont {Dresner}(1995)}]{Dresner2002}%
	  \BibitemOpen
	  \bibfield  {author} {\bibinfo {author} {\bibfnamefont {L.}~\bibnamefont
	  {Dresner}},\ }\href@noop {} {\emph {\bibinfo {title} {{Stability of
	  Superconductors}}}},\ \bibinfo  {editor} {edited by S. Wolf},\ (\bibinfo  {publisher} {Plenum Press},\ \bibinfo
	  {address} {New-York},\ \bibinfo {year} {1995})\BibitemShut {NoStop}%
	\bibitem [{\citenamefont {Levin}\ \emph {et~al.}(2007)\citenamefont {Levin},
	  \citenamefont {Barnes},\ and\ \citenamefont {Bulmer}}]{Levin2007b}%
	  \BibitemOpen
	  \bibfield  {author} {\bibinfo {author} {\bibfnamefont {G.~A.}\ \bibnamefont
	  {Levin}}, \bibinfo {author} {\bibfnamefont {P.~N.}\ \bibnamefont {Barnes}}, \
	  and\ \bibinfo {author} {\bibfnamefont {J.~S.}\ \bibnamefont {Bulmer}},\
	  }\href {\doibase 10.1088/0953-2048/20/8/006} {\bibfield  {journal} {\bibinfo
	  {journal} {Superconductor Science and Technology}\ }\textbf {\bibinfo
	  {volume} {20}},\ \bibinfo {pages} {757} (\bibinfo {year} {2007})}\BibitemShut
	  {NoStop}%
	\end{thebibliography}

\end{document}